\documentclass[aps,prb,twocolumn,superscriptaddress,floatfix,citeautoscript]{revtex4}
\usepackage{times}
\usepackage{graphicx}
\usepackage{dcolumn}
\usepackage{bm}
\usepackage{color}
\usepackage{amssymb}   
\begin{document}

\title{Effective pairing theory for strongly correlated d-wave superconductors}

\author{Debmalya Chakraborty}
\affiliation{Indian Institute of Science Education and Research-Kolkata, Mohanpur, India-741246}
\affiliation{Institut de Physique Th\'eorique, CEA, Universit\'e Paris-Saclay, Saclay, France}\thanks{Present address}

\author{Nitin Kaushal}
\affiliation{Indian Institute of Science Education and Research-Kolkata, Mohanpur, India-741246}
\affiliation{Department of Physics and Astronomy, The University of Tennessee, Knoxville, TN 37996, USA}

\author{Amit Ghosal}
\affiliation{Indian Institute of Science Education and Research-Kolkata, Mohanpur, India-741246}

\begin{abstract}
Motivated by recent proposals of correlation induced insensitivity of d-wave superconductors to impurities, we develop a simple pairing theory for these systems for up to a moderate strength of disorder. Our description implements the key ideas of Anderson, originally proposed for disordered s-wave superconductors, but in addition takes care of the inherent strong electronic repulsion in these compounds, as well as disorder induced inhomogeneities. We first obtain the self-consistent one-particle states, that capture the effects of disorder exactly, and strong correlations using Gutzwiller approximation. These `normal states', representing the interplay of strong correlations and disorder, when coupled through pairing attractions
following the path of Bardeen-Cooper-Schrieffer (BCS), produce results nearly identical to those from a more sophisticated Gutzwiller augmented Bogoliubov-de Gennes analysis.
\end{abstract}

\maketitle

\section{Introduction}\label{sec:Intro}
One of the outstanding puzzles of the disordered superconductors is the insensitivity of the high temperature cuprate superconductors to weak and moderate disorder~\cite{RevModPhys.81.45,Dagotto257,KeimerNature,PhysRevLett.77.5421,2008NatPh...4..762G}. In contrast, the conventional wisdom developed along the lines of Abrikosov-Gorkov (AG) theory~\cite{AGtheory} predicts an extreme sensitivity of these materials to impurities. The original idea, based on perturbative expansions, had been refined subsequently, leading to self-consistent T-matrix calculations~\cite{HIRSCHFELD1986111,PhysRevLett.57.2575,Joynt1997,doi:10.1080/00018730210164638,RevModPhys.78.373,PhysRevB.63.054502}, but the broad sensitivity~\cite{Won2001} of these materials to disorder survived. 

The effects of dopant disorder~\cite{Slezak04032008}, however, on cuprates have remained rather benign.
The inhomogeneities in local doping of the charge carrier induce local variations in the gap map seen from the scanning tunneling microscopy measurements~\cite{PhysRevLett.94.197005,2001Natur.413..282P,2002Natur.415..412L}. Surprisingly, these nanoscale inhomogeneities do not affect the low energy density of states -- as if, the d-wave nodes are ``quantum protected"~\cite{Anderson480}.
The superfluid density and $T_c$ undergo only a modest reductions~\cite{PhysRevB.79.104502,PhysRevLett.91.047001,PhysRevB.53.12454} in spite of the d-wave nature of the anisotropic order parameter~\cite{PhysRevLett.74.797,PhysRevLett.71.2134,PhysRevB.54.R9678,PhysRevLett.73.593}. 
Other unconventional superconductors, e.g. organics~\cite{PhysRevLett.96.177002} and pnictides~\cite{PhysRevB.85.214509,PhysRevLett.117.257002}, which belong to the intermediate coupling category, also feature anomalies.
On the other hand, addition of strong substitutional impurities~\cite{PhysRevLett.77.5421,PhysRevB.56.6201} in these materials weakens superconducting correlations significantly.

A number of non-BCS features of high $T_c$ cuprate superconductors~\cite{RevModPhys.66.763,0953-8984-16-24-R02,RevModPhys.78.17} make them deviate from a favorable playground of AG-type theories. These include the presence of strong repulsive correlations between the charge carriers, short coherence lengths, $\xi$, non-monotonic dependence of $T_c$ on the doping level, small superfluid density etc. In addition, neglect of the spatial fluctuations in the pairing amplitude in a disordered environment in AG formalism calls for a careful microscopic relook into the role of impurities on these systems.

Inclusion of the spatial inhomogeneities of the pairing amplitude for short-coherence length d-wave BCS superconductors (dSC) within a Bogoliubov-de Gennes (BdG) formalism indeed enhances the robustness of dSC to impurities~\cite{PhysRevLett.85.3922,Ghosal00,269e2e1aae614dec84a5de769259f71f,PhysRevB.56.7882}. Recent advances of incorporating the effects of strong electronic repulsions on top of the inhomogeneous background resulted in a Gutzwiller-renormalized theory~\cite{0953-2048-1-1-009,2008NatPh...4..762G,PhysRevB.84.184511,PhysRevB.79.184510}  (referred to as GIMT). These analyses make these superconductors amazingly immune to disorder, up to its strength as large as the bandwidth~\cite{1367-2630-16-10-103018}! Such remarkable robustness of the superconducting correlations~\cite{2008NatPh...4..762G,PhysRevLett.100.257003,1367-2630-16-10-103018,Vlad16}
naturally implies a similar robustness of $T_c$, at least within the mean-field description of the renormalized theory. This raises an intrinsic question: Does Anderson's theorem~\cite{ANDERSON195926}, or an equivalent, apply even for these strongly correlated d-wave superconductors?

We address this question by exploring the fate of a simple-minded pairing theory following Anderson's original idea of `pairing of exact eigenstates'~\cite{ANDERSON195926,PhysRevB.65.014501}. But we upgrade it now to include the inherent strong correlations in these systems, as well as the exact treatment of disorder induced inhomogeneities in our numerical calculations.
It is well established that the `pairing of exact eigenstates' leads to Anderson's theorem for s-wave superconductors (sSC) for weak disorder. However, the same ideas had been successfully extended to incorporate details of inhomogeneities and localization effects in its numerical implementation (for sSC)~\cite{PhysRevB.65.014501}. Here, we expand it further by implementing similar concepts for strongly correlated superconductors with an anisotropic order parameter.

At the outset, we emphasize that our developments pertain to dSC with impurities up to a moderate strength and exclude strong substitutional scatterers. Studies of dSC with strong substitutional impurities, in the limit of unitary scatterers are also available~\cite{PhysRevB.56.6201,PhysRevLett.77.5421,PhysRevLett.91.047001,PhysRevB.51.15547,Hirschfeld15}. There are subtleties in handling strong correlations and also strong impurities in a mean-field formalism~\cite{PhysRevB.95.014516}, and the results depend crucially on their relative strengths.

In this article, we demonstrate that the complexity of strongly correlated disordered superconductors, such as cuprates, can be understood in terms of a simple pairing theory. However, the true potential of our developments lies in identifying the underlying {\it effective} one-particle states, which we termed as `normal states' (NS). It is these states which participate in Cooper-pairing in these materials following the standard BCS path~\cite{PhysRev.106.162}.
We posit that the properties of the true normal state dictate the response of anisotropic superconductors to impurities, providing a deeper insight to the physics of strongly correlated unconventional superconductors.

\section{Model and Methods}\label{sec:model}

\subsection{Anderson's prescription}\label{sec:AP}

The original proposition of the pairing of exact eigenstates, that leads to Anderson's theorem, relies on two important conceptual ideas: (a) The problem of {\it non-interacting} electrons in disorder potential is solved at the first stage to generate its `exact eigenstates'.
BCS type attractive pairing interactions then couple specific pairs of these states producing Cooper-pairs at the second stage and phase coherence of these pairs produce superconductivity in the disordered background. We emphasize that such decoupling of these two stages in the above mechanism necessarily demands that the pairing interactions has no role in determining the exact eigenstates. (b) The specific states participating in Cooper-pairing (at the second stage) are the time reversed exact eigenstates derived in the first stage. This is simply motivated by the BCS theory, which Anderson's pairing method must reduce to, in the clean limit.

Each of these two points are important for establishing Anderson's theorem for disordered sSC. Can they work for the strongly correlated d-wave superconductors as well? In order to explore this question we first set up the formalism below.

\subsection{Normal states: the equivalent of ``exact eigenstates" for strongly correlated dSC}\label{sec:NS}

In the limit when the electron-electron repulsion is strong, it is believed that the phases of the strongly correlated cuprates can be well described by the $``t-J"$ model~\cite{ANDERSON1196}:
\begin{eqnarray}
&&{\cal H}_{\rm t-J} = \sum_{ij \sigma}t_{ij} (\tilde{c}^{\dagger}_{i \sigma} \tilde{c}_{j \sigma}+{\rm h.c.})+\sum_{ij}J_{ij}\Big(\tilde{\mathbf{S}}_i.\tilde{\mathbf{S}}_j-\frac{\tilde{n}_i\tilde{n}_j}{4}\Big) \nonumber \\ 
\label{eq:tJ}
\end{eqnarray}
The first term indicates hopping of electrons on a $2D$ square lattice of $N$ sites. Here, $J$ is the exchange interaction, assumed to arise from a Hubbard-type~\cite{Hubbard238} onsite repulsion $U$ via Schrieffer-Wolff transformation~\cite{0022-3719-10-10-002}, yielding $J_{ij}=4t_{ij}^2/U$. 
We take $t_{ij}=-t$, when $i$ and $j$ are nearest neighbors, denoted as $\langle ij \rangle$, and  $t_{ij}=t'$, when $i$ and $j$ are next-nearest neighbors, with the notation of $\langle \langle ij\rangle \rangle$. We choose $t_{ij}=0$ for all other pairs of $i$ and $j$. Correspondingly, we have $J_{ij}=J$ for $\langle ij \rangle$, $J_{ij}=J'$ for $\langle \langle ij \rangle \rangle$.
Here, $\tilde{c}_{i\sigma}=c_{i \sigma}(1-n_{i\bar{\sigma}})$ is the electron annihilation operator in the `projected Hilbert space' that prohibits double-occupancy at any site $i$, and similarly for the electron creation operator. We introduce disorder by redefining ${\cal H}_{\rm t-J}$ to ${\cal H}_{\rm t-J}+ \sum_{i\sigma} (V_i-\mu)n_{i\sigma}$, where $\mu$ is the chemical potential that fixes the average density of electrons, $\rho = N^{-1}\sum_{i \sigma} \langle n_{i \sigma} \rangle$, in the system to a desired value.
Such a simple re-definition of the Hamiltonian upon inclusion of disorder, however, would not work for strong disorder ($V \ge~3t$) and a modified treatment of Schrieffer-Wolff transformation \cite{PhysRevB.95.014516,RajdeepAbisekh} is necessary.
Here, we use the model of {\it Box-disorder}, where $V_i$'s on all sites $i$ of the lattice are drawn from a uniform `box' distribution, such that, $V_i \in [-V/2,V/2]$ uniformly, thus defining $V$ as the strength of disorder.

We studied the Hamiltonian ${\cal H}_{\rm t-J}$ at zero temperature ($T=0$), upon including disorder, over a wide range of parameters. Here we present results for $U=12t$ and $t'=t/4$ \cite{PhysRevB.52.615}, and we express all energies in the units of $t$. We choose the average density of electrons, $\rho=0.8$, which coincides with the optimal doping. It is the optimal doping where dSC is the strongest in a typical phase diagram of cuprates, in addition to being reasonably free from the complex effects of other competing orders~\cite{Emery03081999,doi:10.1146/annurev-conmatphys-070909-103925,PhysRevB.80.035117,Mesaros08112016,PhysRevB.95.115127,PhysRevB.84.184511,PhysRevLett.99.147002}.
While the phenomenology of competing orders attract interesting and  active research in the underdoped regime \cite{RevModPhys.87.457,PhysRevB.95.104510,PhysRevB.63.094503,PhysRevB.55.14554,PhysRevLett.96.197001}, our goal here is to focus only on the interplay of impurities and strongly correlated dSC, and hence we choose the optimal doping for our study.
We carry out our numerical simulations typically on a $30 \times 30$ lattice, and we collect statistics on our results for each disorder strength $V$ from $10-15$ independent realizations of disorder.

The Hilbert space restriction, that prohibits any double occupancy in the limit of strong correlations, are reflected in the transformation: $c_{i \sigma} \rightarrow \tilde{c}_{i \sigma}$, and makes it difficult to handle these creation and annihilation operators in the projected space. To make progress, we use Gutzwiller approximation (GA)~\cite{0953-2048-1-1-009,doi:10.1080/00018730701627707,PhysRevB.76.245113} to implement the phase space restrictions. GA amounts to renormalizing the parameters $t$ and $J$ of ${\cal H}_{\rm t-J}$ locally by density-dependent factors, such that, they mimic the projection due to strong repulsions. For example, the restricted hopping reduces $t_{ij}$ due to double-occupancy prohibition, whereas, the effective $J_{ij}$ increases because of enhanced overall single-occupancy.
The real advantage of GA lies in the fact that it turns the problem into an effective weak-coupling one redefined in the unprojected Hilbert space, which is now amenable to simple mean field treatments. It has been shown that GA is capable of describing non-BCS and non-trivial features of cuprate superconductors~\cite{PhysRevB.70.054504,PhysRevLett.98.027004} in the clean limit.

Upon carrying out the inhomogeneous Hartree-Fock mean field decoupling of the Gutzwiller renormalized ${\cal H}_{\rm t-J}$ such that {\it no symmetry of ${\cal H}_{\rm t-J}$ is broken}, we arrive at the following `normal state' Hamiltonian:
   \begin{eqnarray}
&&{\cal H}_{\rm NS}  = \sum_{i, \delta, \sigma}{} \{ t_{i\delta} g^t_{i,i+\delta}-W^{\rm FS}_{i\delta} \} ~c^{\dagger}_{i \sigma} c_{i+\delta \sigma} \nonumber \\
&+& \sum_{i, \tilde \delta, \sigma}{} \{ t_{i \tilde \delta} g^t_{i,i+\tilde \delta} \} ~c^{\dagger}_{i \sigma} c_{i+\tilde \delta \sigma}+\sum_{i,\sigma} (V_i-\mu+\mu_i^{\rm HS}) n_{i\sigma} \nonumber \\
\label{eq:meanfield1}
\end{eqnarray}
Here, we have written the Hamiltonian on bonds connecting sites $i$ and $j$, where $j=i+ \delta$, with $\delta=\pm x ~{\rm or} \pm y$, $\tilde{\delta}=\pm (x \pm y)$. As the name suggests, we refer to the eigenstates of ${\cal H}_{\rm NS}$ in Eq.~(\ref{eq:meanfield1}) as the {\rm normal states}, ${\rm NS}_{\rm GIMT}$ (here ${\rm GIMT}$ in the subscript of normal states, NS, stands for Gutzwiller-augmented inhomogeneous Hartree-Fock mean-field theory).
It is crucial to include the effect of strong correlations in ${\cal H}_{\rm NS}$ following the above construction, even though the final one-particle Hamiltonian without broken symmetries is similar in structure to the disordered tight binding model (or Anderson model of disorder). Yet, these normal states distinguish themselves from the `exact eigenstates' (eigenstates of the Anderson model of disorder) in accounting for the strong correlation effects through Gutzwiller factors, as well as the Hartree- and Fock-shifts. These considerations naturally make the solution of ${\cal H}_{\rm NS}$ already a self-consistent problem. We also emphasize that these normal states are defined at $T=0$, and are not to be confused with the common notion of the high temperature normal state of the material in which thermal fluctuations destroy superconductivity. The Fock-shift ($W^{\rm FS}_{i\delta}$) and the Hartree-shift ($\mu_i^{\rm HS}$) terms in Eq.~\ref{eq:meanfield1} are given by, 
\begin{equation}
W_{i \delta}^{FS}=\frac{J}{2} \left\{ \Bigg( \frac{3g_{i,i+\delta}^{xy}}{2} -\frac{1}{2} \Bigg)  \tau_{i}^{\delta} \right\}
\label{eq:fockshift}
\end{equation}
\begin{eqnarray}
\mu_i^{HS}&=&-4t\sum_{\delta,\sigma} \left\{ \frac{\partial{g_{i,i+\delta}^{t}}}{\partial{\rho_i}} \tau_i^{\delta} \right\}+4t'\sum_{\tilde{\delta},\sigma} \left\{ \frac{\partial{g_{i,i+\tilde{\delta}}^{t}}}{\partial{\rho_i}} \tau_i^{\tilde{\delta}} \right\} \nonumber \\
&-&\frac{3J}{2}\sum_{\delta,\sigma} \frac{\partial{g_{i,i+\delta}^{xy}}}{\partial{\rho_i}}\Big( {\tau_i^{\delta}}^2 \Big) 
\label{eq:hartreeshift}
\end{eqnarray}
where, $\rho_i= \sum_{\sigma} \langle n_{i \sigma} \rangle_0$ and $\tau_{ij} \equiv \langle c_{i \downarrow}^{\dagger} c_{j \downarrow} \rangle_0 \equiv \langle c_{i \uparrow}^{\dagger} c_{j \uparrow} \rangle_0$. Here, $\langle \rangle_0$ denotes the expectation value in the unprojected space. The Gutzwiller factors in Eq.~(\ref{eq:fockshift}) and (\ref{eq:hartreeshift}) are given in terms of the local density:
\begin{equation}
g^t_{ij}=\sqrt{\frac{4(1-\rho_i)(1-\rho_j)}{(2-\rho_i)(2-\rho_j)}}, ~~g^{xy}_{ij}=\frac{4}{(2-\rho_i)(2-\rho_j)}
\label{eq:gut1}
\end{equation}
As mentioned, the above construction of the ${\rm NS}_{\rm GIMT}$ excludes any broken symmetry order parameters, e.g. magnetism, charge density wave etc. However, unbroken symmetry is not a fundamental requirement of ${\rm NS}_{\rm GIMT}$. In fact, we need to include them in ${\cal H}_{\rm NS}$ (except, of course, any superconducting order through Bogoliubov channels), when we study the effects of such additional orders competing with superconductivity. 

Considering the unitary transformation to diagonalize ${\cal H}_{\rm NS}$ in the $\{\alpha\}$-basis:
\begin{equation}
{c}_{i \sigma} = \sum_{\alpha =1}^N \psi_i^\alpha c_{\alpha \sigma},
\label{eq:nstrans}
\end{equation}
we obtain ${\cal H}_{\rm NS} = \sum_{\alpha,\sigma}{\xi}_{\alpha}c_{\alpha \sigma}^{\dagger}c_{\alpha \sigma}$. Here, the self-consistent $\{\psi_i^\alpha\}$ are the eigenvectors of ${\cal H}_{\rm NS}$, and they constitute our ``normal states", i.e. the ${\rm NS}_{\rm GIMT}$.

\subsection{Pairing of normal States (PNS)}\label{sec:PNS}

To study the superconducting properties of ${\cal H}_{\rm t-J}$ in Eq.~(\ref{eq:tJ}), we now introduce the pairing term,
\begin{equation}
{\cal H}_{\rm P} = \frac{1}{2} \sum_{\langle ij \rangle} \Delta_{ij}(c_{i\uparrow}^{\dagger}c_{j\downarrow}^{\dagger} - c_{i\downarrow}^{\dagger}c_{j\uparrow}^{\dagger}) + {\rm h.c.}
\label{eq:Hpar}
\end{equation}
in addition to ${\cal H}_{\rm NS}$, where, 
\begin{equation}
\Delta_{ij}=  -\frac{J}{2}\Big(\frac{3g^{xy}_{ij}+1}{4}\Big) { \langle  c_{i\uparrow}c_{j\downarrow}\rangle_0 - \langle  c_{i\downarrow}c_{j\uparrow}\rangle_0. }  
\label{eq:deltadef}
\end{equation}
The pairing part of the Hamiltonian in Eq.~(\ref{eq:Hpar}) can be thought to arise from a mean-field decoupling of the original ${\cal H}_{\rm t-J}$ in the Bogoliubov channel. Note that, the form of ${\cal H}_{\rm P}$ ensures that we have chosen the singlet pairing channel on the links.
Writing $H_{\mathcal P}$ in the $\{\alpha\}$-basis, we have,
\begin{equation}
{\cal H}_{\rm P}= \frac{1}{2} \sum_{\alpha \beta} \Delta_{\alpha \beta}\{  c_{\alpha\uparrow}^{\dagger} c_{\beta\downarrow}^{\dagger} -  c_{\alpha\downarrow}^{\dagger} c_{\beta\uparrow}^{\dagger}  \}  + {\rm h.c.}
\label{eq:pareigen}
\end{equation}
where, 
\begin{equation}
\Delta_{\alpha \beta} = \sum_{\langle ij \rangle}\Delta_{ij}(\psi_i^{\alpha})^*(\psi_j^\beta)^*,
\label{eq:deleig}
\end{equation}
leaving the total Hamiltonian as:
\begin{eqnarray}
{\cal H}_{\rm total}&=& \sum_{\alpha,\sigma} ({\xi}_{\alpha}- \mu_{p}) c_{\alpha \sigma}^{\dagger}c_{\alpha \sigma} \nonumber \\ 
&+& \frac{1}{2} \sum_{\alpha \beta} \Big( \Delta_{\alpha \beta}\{   c_{\alpha\uparrow}^{\dagger} c_{\beta\downarrow}^{\dagger} -  c_{\alpha\downarrow}^{\dagger} c_{\beta\uparrow}^{\dagger}  \}  + {\rm h.c.} \Big) 
\label{eq:totalhamil}
\end{eqnarray}
Here, we introduced $\mu_{p}$ to fix the final average density (after pairing) to the desired value $\rho=0.8$. Note that, the $\mu$ in ${\cal H}_{\rm NS}$ fixes the density to the same desired value, but only in the normal state. Pairing at the second stage (after inclusion of ${\cal H}_{\rm P}$) can deviate $\rho$ from this value. We use $\mu_{p}$ to tune it back to the chosen value.
We also note that there is no restriction, in principle, on $\alpha$, $\beta$ in the definition of $\Delta_{\alpha \beta}$ appearing in Eq.~(\ref{eq:deleig}), though we will see in Sec. (\ref{sec:pwidth}) that the dominant contribution comes from those {$\alpha$, $\beta$} for which $\xi_\alpha \approx \xi_\beta$.

\subsection{Self-consistent pairing amplitude}\label{sec:selfcon}

Evidently, ${\cal H}_{\rm total}$ in Eq.~(\ref{eq:totalhamil}) carries the BCS structure and we diagonalize it using a modified Bogoliubov transformation:
\begin{equation}
c_{p\sigma} = \sum_{n=1}^{N} \left( u_{p,n} \gamma_{n\sigma} - \sigma v_{p,n}^* \gamma_{n\bar {\sigma}}^{\dagger} \right)
\label{eq:btypetrans}
\end{equation}
where, $\gamma_{n {\sigma}}^{\dagger}$ ($\gamma_{n\sigma}$) are fermionic quasiparticle creation (annihilation) operators.

Starting with guess values of $\Delta_{ij}$ on all the $2N$ bonds we first obtain the $N^2$ numbers of $\Delta_{\alpha \beta}$ using the normal state eigenfunctions $\psi^{\alpha}_i$'s in Eq.~(\ref{eq:deleig}). The eigenvalues and eigenvectors of ${\cal H}_{\rm total}$ allows us to re-calculate $\Delta_{ij}$ and $\rho_i$ using Eq.~(\ref{eq:deltadef}) and the self consistency conditions:
\begin{equation}
\langle  c_{i\uparrow}c_{j\downarrow}\rangle_0 = \sum_{p_1,p_2=1}^{N}\psi_i^{p_1}\psi_j^{p_2}\langle  c_{p_1\uparrow} c_{p_2\downarrow}\rangle
\label{eq:selfcs1}
\end{equation}
\begin{equation}
\rho_{i} =  2 \sum_{p_1,p_2=1}^{N}(\psi_i^{p_1})^* \psi_{i}^{p_2}\langle c_{p_1\downarrow}^{\dagger} c_{p_2\downarrow} \rangle.
\label{eq:selfcs2}
\end{equation}
We then iteratively update the guess values of $\Delta_{ij}$ and $\rho_i$ for the inputs in Eq.~(\ref{eq:totalhamil}) in order to achieve the final self-consistency until the inputs and corresponding outputs in Eq.~(\ref{eq:selfcs1}) and (\ref{eq:selfcs2}) match within tolerance. For accelerating the convergence, we used combinations of linear, Broyden and modified Broyden~\cite{PhysRevB.38.12807} schemes of mixing of the input and output at every iteration. 

\begin{figure}[t]
\centering
  \begin{tabular}{@{}cc@{}}
    \includegraphics[width=.198\textwidth]{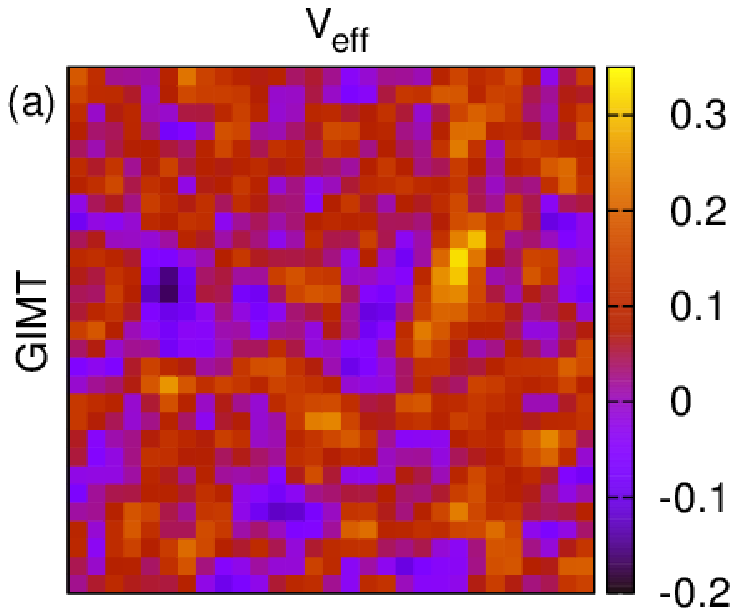} &
    \includegraphics[width=.202\textwidth]{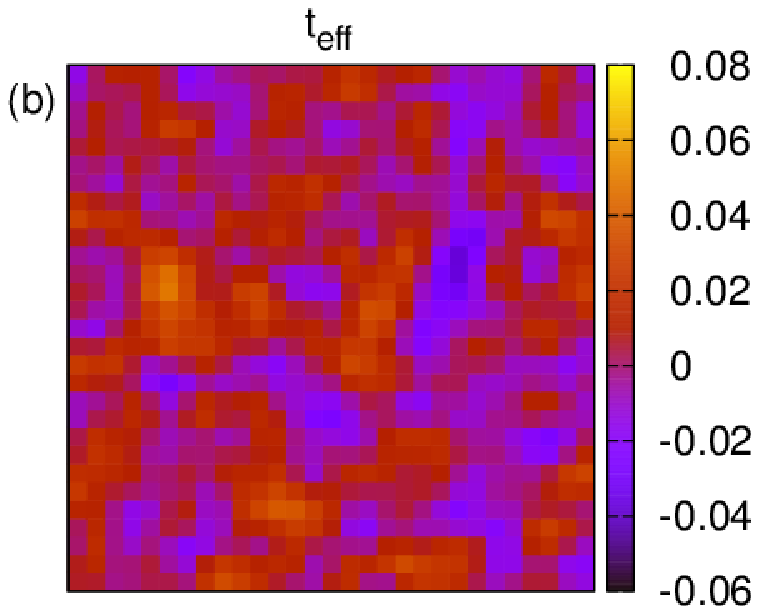} \\
    \includegraphics[width=.198\textwidth]{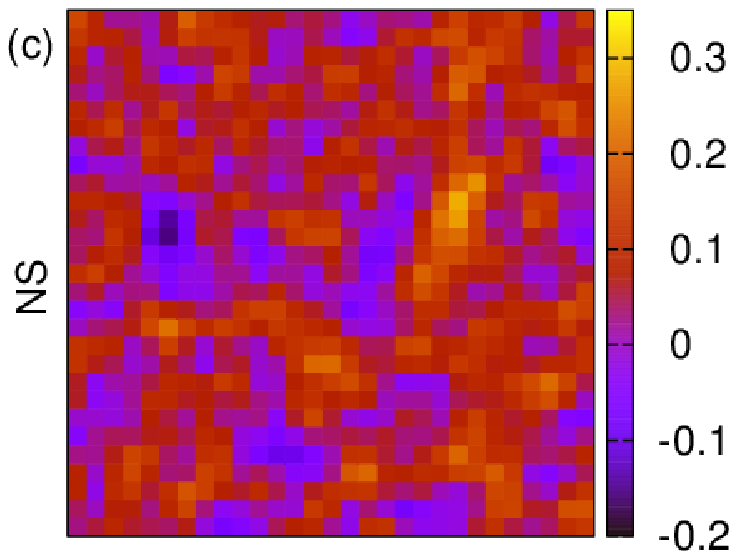} &
    \includegraphics[width=.202\textwidth]{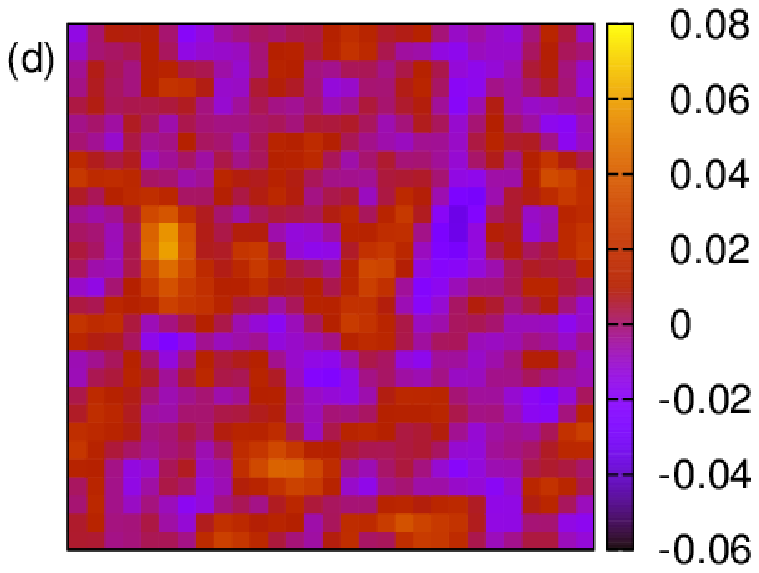} \\
    \includegraphics[width=3.8cm,height=3.2cm]{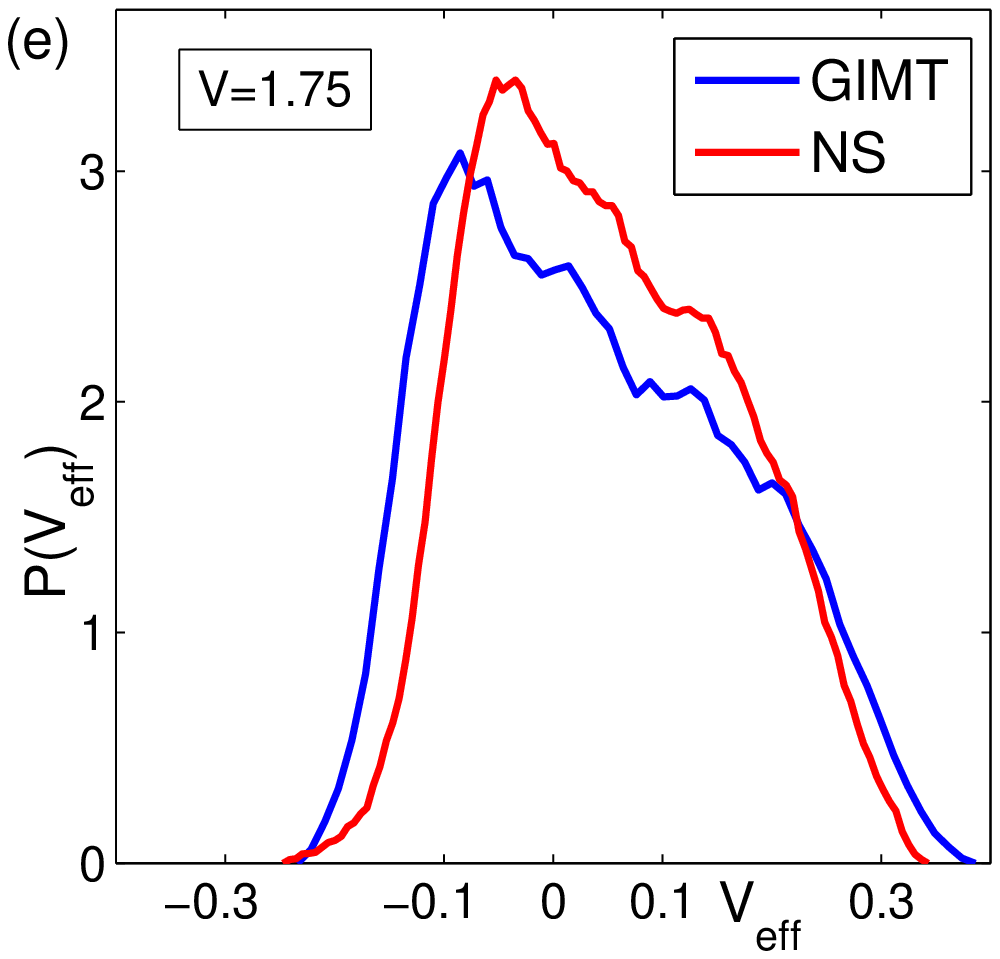} &
    \includegraphics[width=3.7cm,height=3.2cm]{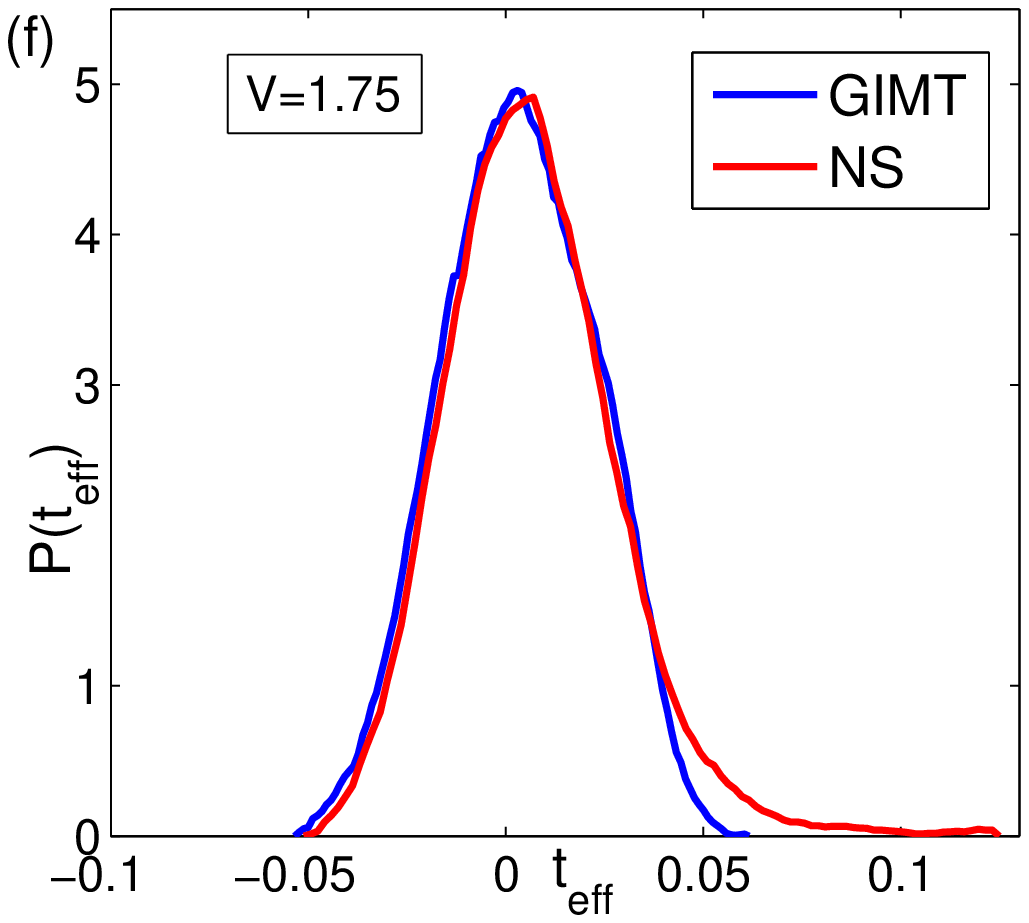} 
  \end{tabular}
  \caption{(a-d) Spatial density map of $V_{\rm eff}$ and $t_{\rm eff}$ from NS and GIMT shows similar spatial anticorrelation between $V_{\rm eff}$ and $t_{\rm eff}$. They also highlight the spatial correlation in $V_{\rm eff}$. Comparison of distributions $P(V_{\rm eff})$ in (e) and $P(t_{\rm eff})$ in (f) for ${\rm NS}$ and for the full GIMT outputs at $V=1.75$. The distributions match rather well in the two calculations validating the basis of PNS calculations.
We subtracted the homogeneous components of $t_{\rm eff}$ ($t_{\rm eff}(V=0)=0.459$) and $V_{\rm eff}$ ($V_{\rm eff}(V=0)=1.6$), arising from the Fock- and Hartree-shift respectively. The resulting distributions in (e) and (f) feature zero mean -- this is broadly true for all $V$.
}
\label{fig:teffveff}
\end{figure}
\section{Results}\label{sec:Res}

We will discuss in this section our findings from the pairing of normal states (PNS) and compare them with GIMT findings. Here, GIMT refers to the full BdG calculation augmented with Gutzwiller renormalization. However, it is truly  illuminating to focus our attention first on the distinguishing features of ${\rm NS}_{\rm GIMT}$ that separate them from their uncorrelated counterparts -- the ``exact eigenstates" of the Anderson's model of disorder. 
\begin{figure}[t]
\includegraphics[width=0.45\textwidth]{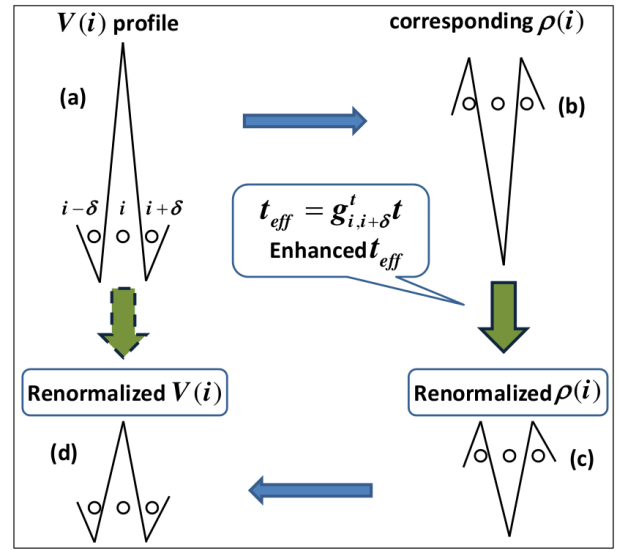}
\caption{ A schematic evolution of the inhomogeneity in space that leads to the renormalization of $V_{\rm eff}(i)$ and spatial anti-correlation between $V_{\rm eff}(i)$ and $t_{\rm eff}(i)$, upon including electronic repulsions through Gutzwiller approximation (GA). Consider in (a) the site $i$ having a high hill of disorder potential (also assumed that $V_{i \pm \delta}=0$), that would normally yield a low $\rho_i$ compared to $\rho_{i \pm \delta} \approx \rho_0$, as shown in (b), rarely populating the site $i$. However, GA insures that $t_{\rm eff}$ on bonds connecting $i$ to its neighbors is enhanced, according to Eq.~(\ref{eq:gut1}), increasing charge flow to this site. This in turn reduces the dip in $\rho_i$ as seen in (c), so that the corresponding $V_{\rm eff}(i)$, that would have normally produced the $\rho_i$ in (c), is far weaker than its bare value, shown in (a). Exactly similar arguments would yield a similar weakening of deep potential well by strong correlations. 
}
\label{fig:schem}
\end{figure}
\subsection{Structure of the {\it normal states}}\label{sec:StrucNS}

For the convenience of our discussions below, it is useful to cast the normal state Hamiltonian ${\cal H}_{\rm NS}$
in the following form:
\begin{equation}
{\cal H}_{\rm NS} = -\sum_{i,\delta,\sigma} t_{\rm eff}(i,\delta)c^{\dagger}_{i,\sigma} c_{i+\delta,\sigma} + \sum_{i,\sigma} V_{\rm eff}(i) n_{i,\sigma},
\label{eq:Hns}
\end{equation}
to emphasize ${\cal H}_{\rm NS}$ as a tight binding model, but with effective disorder {\it both} on the links ($t_{\rm eff}$), as well as on the sites ($V_{\rm eff}$). However, these disorder terms now contain order parameters, as seen from Eq.~(\ref{eq:meanfield1}), (\ref{eq:fockshift}), and (\ref{eq:hartreeshift}), and hence, must be evaluated self-consistently, as mentioned already. We find them to develop spatially correlated structures, and are illustrated in Fig.~(\ref{fig:teffveff} a-d). For a justified comparison between the spatial structures of $V_{\rm eff}$ and $t_{\rm eff}$, we transformed the bond variable $t_{\rm eff}(i,\delta)$ to a site variable using relation: $t_{\rm eff}(i)=\frac{1}{4}\sum_{\delta} t_{\rm eff}(i,\delta)$. Spatial associations are found, firstly, in the profile of $V_{\rm eff}(i)$ itself, showing conglomeration of regions with large and small $V_{\rm eff}$, but more importantly, through the explicit anti-correlation of regions of $V_{\rm eff}$ and $t_{\rm eff}$ in space.
We also compare the distributions $P(V_{\rm eff})$ and $P(t_{\rm eff})$ for $V=1.75$ from the ${\rm NS}_{\rm GIMT}$ and GIMT results in Fig.~(\ref{fig:teffveff} e,f), using statistics over $10$ realizations of disorder. Such a favorable comparison of ${\rm NS}_{\rm GIMT}$ outputs of $t_{\rm eff}$ and $V_{\rm eff}$ with those from GIMT validates the conceptual basis of the PNS formalism.
The role of strong electronic repulsions on the disordered normal states has a simple and intuitive rationale, as we describe below. Random impurity potential tends to generate charge inhomogeneities in space, whereas, repulsive interactions smear out such heterogeneities, trying to restore its homogeneous distribution. The key ingredient of ${\rm NS}_{\rm GIMT}$, that distinguishes it from the `exact eigenstates', lies in its impurity renormalization -- a footprint of electronic repulsion in ${\rm NS}_{\rm GIMT}$. This is ascribed to the modification the hopping amplitudes based on local density, which smear out charge accumulation near deep potential wells, and also partly populating potential hills, as explained in Fig.~(\ref{fig:schem}).
As a schematic description, we consider in Fig.~(\ref{fig:schem}), a site $i$ having a high hill of local potential, and hence it ordinarily supports little density of electrons there, compared to the average density on its neighbors, assumed to have no disorder. This local charge imbalance leads to an interesting feedback loop through $g^{t}_{ij}$, absent in the uncorrelated systems. The low electronic density at $i$ enhances $g^{t}_{ij}$ according to Eq.~(\ref{eq:gut1}), which in turn enhances the charge fluctuations across site $i$, leading to an larger effective $\rho_i$ than what would be its value in the absence of the Gutzwiller factors. This leads to a much weaker effective disorder~\cite{PhysRevB.79.184510,PhysRevB.91.020501,1367-2630-16-10-103018,PhysRevLett.91.066603} to account for the enhanced $\rho_i$. In addition to impurity renormalization, the above argument sheds light on  the spatial anti-correlations of $V_{\rm eff}$ and $t_{\rm eff}$. Both these features make ${\rm NS}_{\rm GIMT}$ distinct from the plain `exact eigenstates'.
However, in the limit $U \rightarrow 0$, the ${\rm NS}_{\rm GIMT}$ and `exact eigenstates' would be identical.
\begin{figure}[t]
\includegraphics[width=0.4\textwidth]{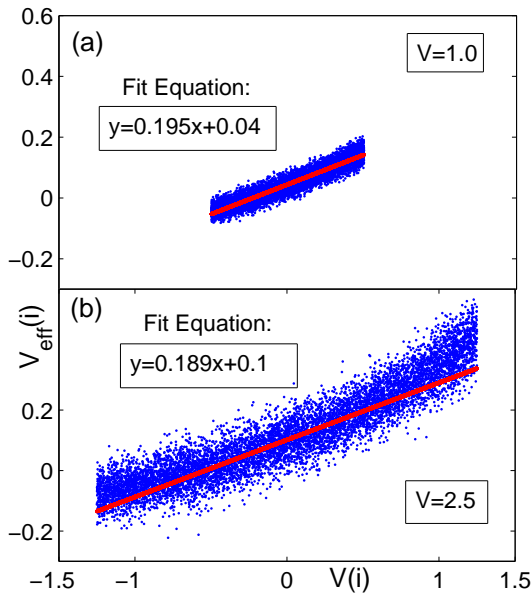}
\caption{Scatter plots of $V_{\rm eff}(i)$ against the bare potential $V(i)$ for: (a) $V=1.0$, and (b) $V=2.5$. The red lines are the best fit to the data. The slope of the solid line in both panels are close to the average doping ($\delta=0.2$). For $V=2.5$, the data tend to deviate from the fit for larger $|V(i)|$, signalling higher order effects.
}
\label{fig:scatter}
\end{figure}
How strong is such renormalization of disorder? In order to get a quantitative estimate of the impurity renormalization, we present the scatter plot of $V_{\rm eff}$ against bare $V$ in Fig~(\ref{fig:scatter}) from our self-consistent NS-calculations (statistics collected over 10 realizations of disorder). Our results show a simple linear trend: $V_{\rm eff} \approx \delta V$ for low $V$, with weak corrections for stronger $V$. Here, $\delta=(1-\rho)$ is the average doping. This low-$V$ linearity is consistent with earlier findings from a single-impurity calculation~\cite{PhysRevB.79.184510}. This is easily comprehended: Since $t \rightarrow g^{t}t \sim \delta t$, we must rescale $V$ by the same factor for a justified comparison, yielding $V_{\rm eff} \sim \delta V$. For the cuprate superconductors, we typically have $\delta \le 0.2$. The above considerations then imply that the Fermi's golden rule estimate of the inverse scattering time of the electrons in the underlying ${\rm NS}_{\rm GIMT}$ is an order of magnitude smaller compared to the `usual' exact eigenstates: $\tau_{\rm NS}^{-1} \sim \tilde{g}(0) V_{\rm eff}^2 \sim \delta \tau_{0}^{-1}$, where $\tilde{g}(0)$ is the density of states at Fermi energy of ${\rm NS}_{\rm GIMT}$. A similar dependence of $\tau^{-1}$ has also been been predicted recently from the T-matrix estimation~\cite{Vlad16}.
We focus next on Cooper-pairing between these strongly correlated ${\rm NS}_{\rm GIMT}$ states. 
\begin{figure}[t]
\includegraphics[width=0.5\textwidth]{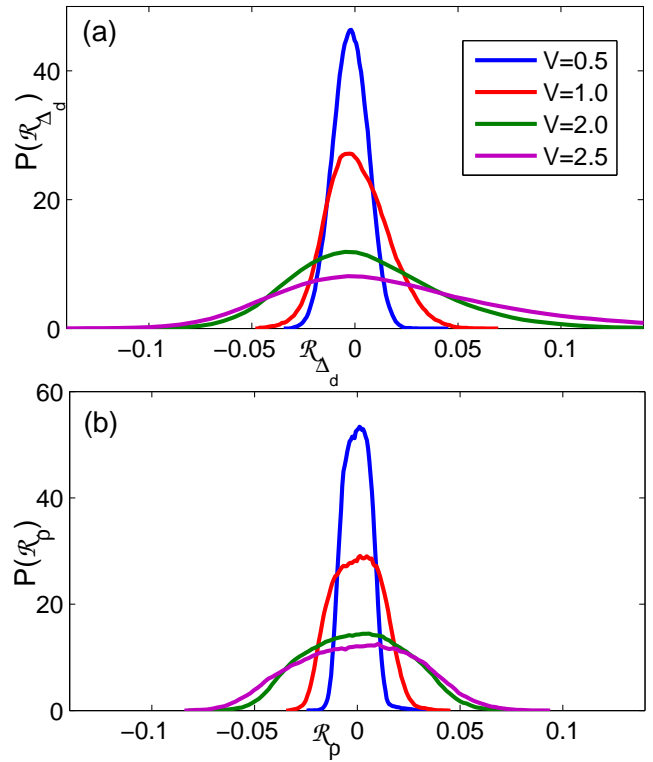}
\caption{The distribution $P({\cal R}_{\Delta_d})$ in (a) and of $P({\cal R}_{\rho})$ in (b) are shown for various disorder strengths. Sharply peaked nature of these distributions (with small variance) validates the PNS formalism.
}
\label{fig:figure1}
\end{figure}
\subsection{Self-consistent order parameters}\label{sec:OP}
 
Inducing pairing through BCS-type attraction as described in Sec,~(\ref{sec:PNS}), we find that the self-consistent PNS outputs of the spatial profiles of the pairing amplitude $\Delta_{ij}$, local density $\rho_i$, or $\tau_{ij}$ are nearly indistinguishable from the results of GIMT calculations.
In order to quantify the strength of PNS formalism, we find it easier to define the relative difference in the PNS order parameters with respect to those from GIMT, in the following manner:
\begin{equation}
{\cal R}_{\rm OP}(i)=\frac{OP^{\rm GIMT}(i)-OP^{\rm PNS}(i)}{\langle OP \rangle ^{\rm GIMT}}
\label{relop}
\end{equation}
where, $OP$ represents either of $\rho_i$, $\Delta_d(i)$ or $\tau_{ij}$. Here, $\langle \rangle$ denotes average over all sites and over configurations. We define the d-wave superconducting order parameter on a site as: $\Delta_d(i)=\frac{1}{4}(\Delta_i^{ +x} - \Delta_i^{ +y} + \Delta_i^{ -x} -\Delta_i^{ -y})$.
We emphasize here that the PNS self-consistency produces for us the solution of link variable $\Delta_{ij}$ among other things. This by itself is no confirmation of a d-wave anisotropy of the pairing amplitude. However, our choice of parameters in the Hamiltonian ${\cal H}_{\rm t-J}$ ensures that we have {\it exclusively} the d-wave ($d_{x^{2}-y^{2}}$) pairing amplitude in the clean limit. Introduction of disorder does generate other possibilities of bond pairing amplitude, e.g., $\Delta_{xs}$, $\Delta_{s_{xy}}$, $\Delta_{d_{xy}}$ \cite{PhysRevLett.101.206404}. But their strengths remain negligibly small compared to the $\Delta_d$ component. GIMT calculations also confirm the same qualitative picture in this regard.

We plot the normalized distribution of ${\cal R}_{\Delta_d}$ and ${\cal R}_{\rho}$ for different $V$ in Fig.~(\ref{fig:figure1}). These distributions, always peaked at zero, show only a weak broadening with $V$. Further, such smearing is essentially independent of $V$ in the range $1.5 \le V \le 2.5$.
The difference between PNS and GIMT remains only at about $3\%$ for all order parameters at $V=2.5$, emphasizing the accuracy of the proposed PNS method to describe the strongly correlated dSC.

\begin{figure}[t]
\includegraphics[width=0.5\textwidth]{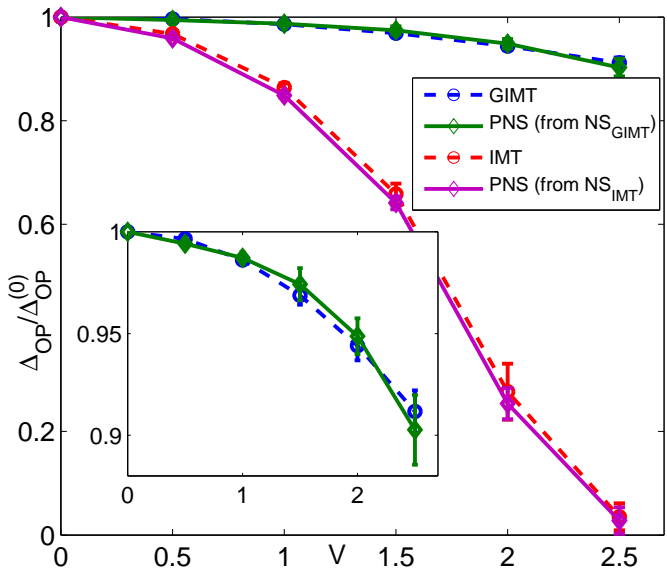}
\caption{Evolution of $\Delta_{\rm OP}$ is presented against $V$. The $V$-dependences of both the PNS and GIMT results show nearly identical behaviour. The inset shows an expanded region of the main panel establishing that the PNS findings match excellently with those from GIMT within the error bars. The results for $\Delta_{\rm OP}$ from IMT calculations, shown by the magenta curve (forcing all Gutzwiller factors to unity, and thereby neglecting strong electronic repulsions), deviate significantly from the PNS or GIMT results. However, it still complements the plain BdG results (red dashed line) exceedingly well. 
}
\label{fig:odlro}
\end{figure}
\subsection{Off-diagonal long range order}\label{sec:odlro}

In order to illustrate the accuracy of the PNS results for physical observables, we study the $V$-dependence of the superconducting off-diagonal long range order (ODLRO), defined as:
\begin{equation}
\Delta_{\rm OP}^2=\lim_{|i-j| \to \infty} F_{\delta, \delta'}(i-j)
\label{eq:odlro}
\end{equation}
where, the pair-pair correlation function, $F_{\delta, \delta'}(i-j)=\langle B_{i \delta}^{\dagger} B_{j \delta'} \rangle$, and, $B_{i \delta}^{\dagger}=( c^{\dagger}_{i \uparrow} c^{\dagger}_{i + \delta \downarrow} + c^{\dagger}_{i + \delta \uparrow} c^{\dagger}_{i  \downarrow})$ is the singlet Cooper-pair creation operator on the links connecting the neighboring sites at $i$ and $i+\delta$.
Since $F_{\delta, \delta'}(i-j)$ can be interpreted as simultaneous hopping of a singlet cooper-pair on a link, the Gutzwiller factor corresponding to this process becomes $g^{t}_{i,j}g^{t}_{i+\delta,j+\delta'}$. We calculate $F_{\delta, \delta'}(i-j)$ using the transformations Eq.~(\ref{eq:nstrans}) and (\ref{eq:btypetrans}). The evolution of ODLRO (normalized by its value $\Delta^{(0)}_{\rm OP}$ at $V=0$) with $V$, as evaluated from the PNS and GIMT calculations, are shown in Fig.~(\ref{fig:odlro}).
The main panel shows that the PNS results are nearly identical with the GIMT findings (see the inset for an expanded view), ascertaining that PNS formalism serves as good a purpose as the GIMT method for handling the physics of strong correlations.

An independent test for the effectiveness of the PNS formalism comes from its comparison with a full BdG calculation, when both neglects strong correlations (and will be referred to as IMT, henceforth). Suppression of strong correlations, though unphysical for cuprates, can easily be implemented by setting all Gutzwiller factors to unity. In Fig.~(\ref{fig:odlro}) we also compared $\Delta_{\rm OP}(V)$ as obtained from pairing between ${\rm NS}_{\rm IMT}$ with those from corresponding plain BdG outcomes. The excellent match of the two formalisms even in the uncorrelated domain strengthens PNS method as a natural description of disordered superconductors. Note that the results differ significantly by including and excluding Gutzwiller factors -- irrespective of PNS or BdG methods (See also Sec.~(\ref{sec:conc})). 

\subsection{Pairing of limited states with close by energies}\label{sec:pwidth}

As discussed in Sec.~(\ref{sec:PNS}), the PNS method amounts to pairing between all the eigenstates of ${\cal H}_{\rm NS}$, making its numerical implementation computationally as demanding as that of GIMT. However, technical gain can be insured by having to pair only a limited number of normal states $\alpha$ and $\beta$ that are not too far from the Fermi energy, such that, $\xi_{\alpha} \approx \xi_{\beta}$. Such an expectation is, of course, motivated by the structure of the BCS gap equation.
\begin{figure}[t]
\centering
  \begin{tabular}{@{}cc@{}}
    \includegraphics[width=3.99cm,height=3.74cm]{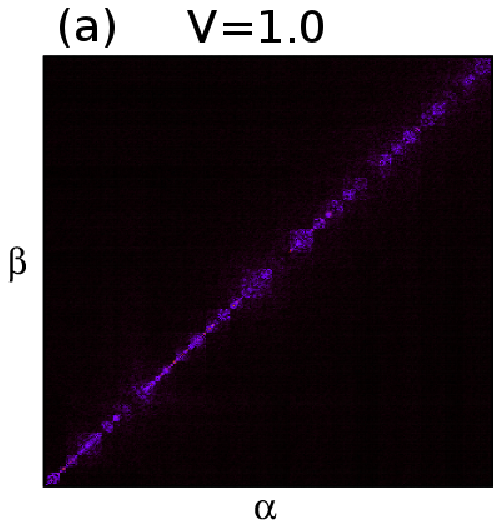} &
    \includegraphics[width=3.99cm,height=3.74cm]{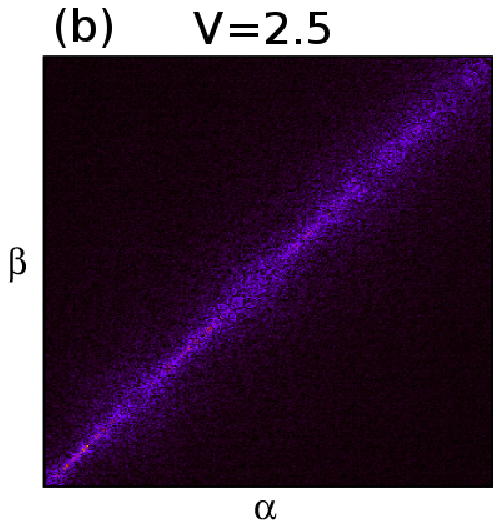} \\
    \includegraphics[width=3.73cm,height=3.74cm]{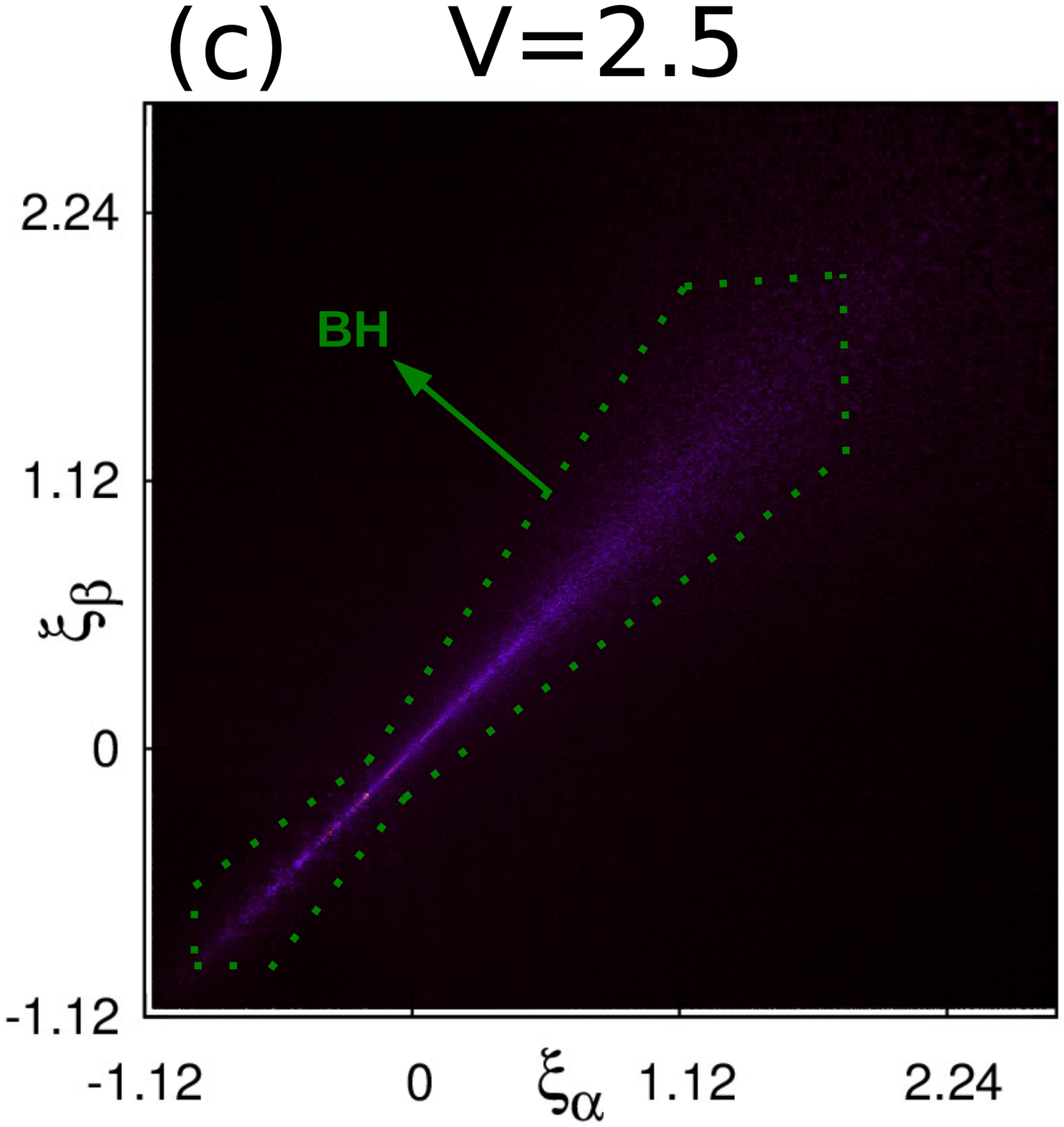} &
    \includegraphics[width=3.73cm,height=3.74cm]{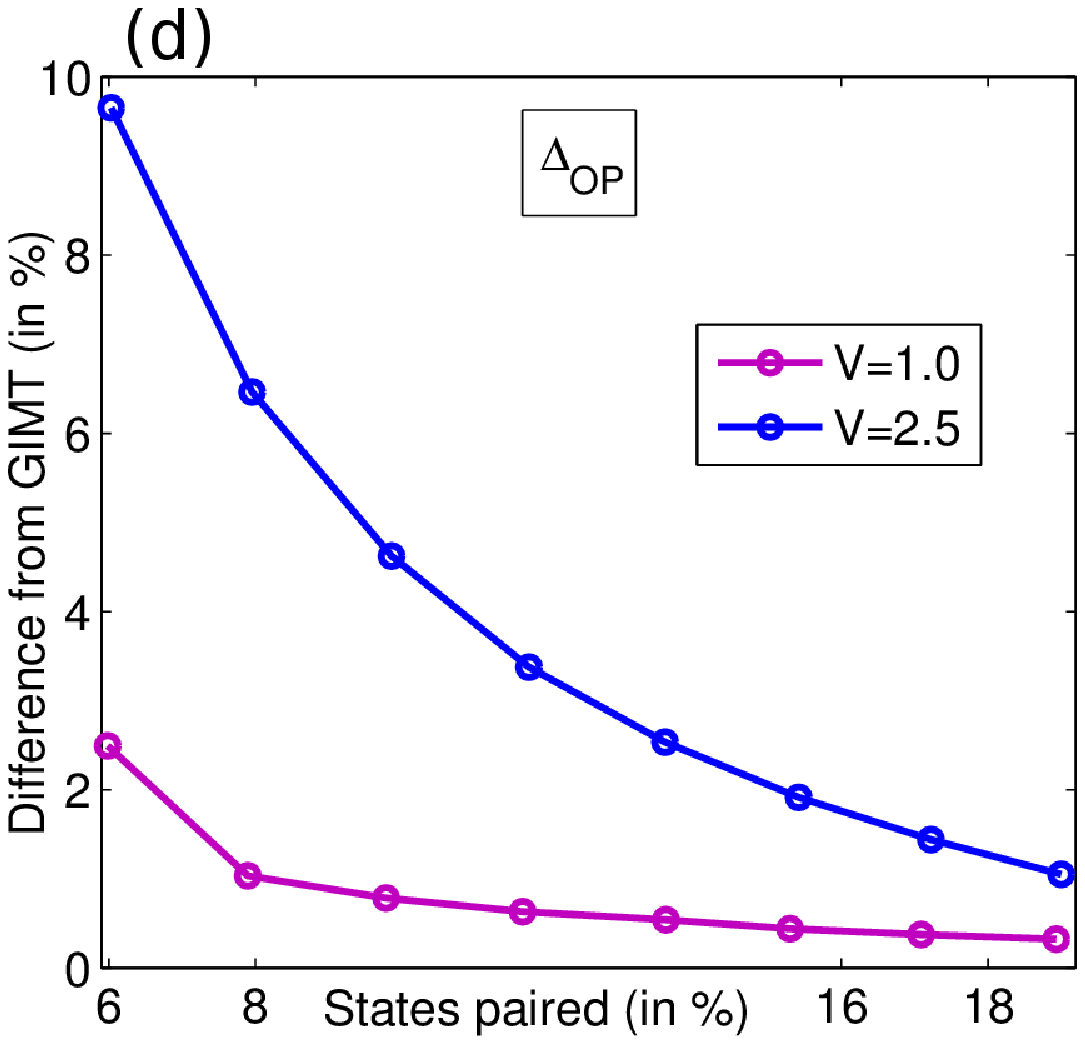} 
  \end{tabular}
  \caption{Intensity plot of $|\Delta_{\alpha \beta}|$ in the normal state eigen basis $\alpha$-$\beta$ for (a) $V=1.0$, and (b) $V=2.5$. We show $|\Delta_{\alpha \beta}|$ in a limited range of $\alpha, \beta$ (only the central part) for a better resolution. The presented values of $|\Delta_{\alpha \beta}|$ are scaled by their maximum values for clarity ($0.35$ for $V=1.0$ and $0.3$ for $V=2.5$). The near-diagonal nature of the pairing is evident for both $V$. The color scales are identical to that in Fig.~(\ref{fig:teffveff}). (c) Density-plot of $|\Delta_{\alpha \beta}|$ against $\xi_{\alpha}$ and $\xi_{\beta}$ across the full (renormalized) bandwidth for $V=2.5$. While the diagonal character of $|\Delta_{\alpha \beta}|$ is evident, only negligible contribution to $|\Delta_{\alpha \alpha}|$ comes from the states near band edges. (d) Accuracy of PNS (with respect to GIMT) is shown along $y$-axis, against the percentage of states paired (along $x$-axis). This accuracy, already impressive with about 10\% ${\rm NS}_{\rm GIMT}$ participating in pairing, becomes better as more states included in Cooper-pairing.}
\label{fig:width}
\end{figure}

In search of this simplification, we plot in Fig.~(\ref{fig:width} a,b), the fully self-consistent and disorder averaged profiles of $|\Delta_{\alpha \beta}|$ in the eigen-space of $\alpha$ and $\beta$. The near diagonal structures of $|\Delta_{\alpha \beta}|$ implies that the states $\alpha$, $\beta$ which are far in energies, have negligible contributions in $\Delta_{\alpha \beta}$. Such a diagonal character of $\Delta_{\alpha \beta}$ is well maintained for $V \le 3$.

The diagonal nature is preserved when the same $|\Delta_{\alpha \beta}|$ is plotted against $\xi_{\alpha}$ and $\xi_{\beta}$ (shown for $V=2.5$ in Fig.~(\ref{fig:width}c)). We also note that not all normal states contribute to `diagonal' pairing, particularly those states lying close to band edges contribute only negligibly to $|\Delta_{\alpha \alpha}|$. Such contribution would have been limited only to a narrow energy window, $\pm \hbar \omega_D$, in simple BCS theory ($\omega_D$ being the Debye frequency). In the present case of strongly correlated anisotropic superconductors in the presence of disorder, the energy range of contribution is wider. Further, the inclusion of next-nearest neighbor hopping, $t'$, makes the ${\rm NS}_{\rm GIMT}$ energy-band (and hence $|\Delta_{\alpha \alpha}|$) asymmetric about the Fermi energy ($\xi=0$). The final profile of $|\Delta_{\alpha \beta}|$, as seen from Fig.~(\ref{fig:width}c), hints that the summations in Eqs.~(\ref{eq:selfcs1}) and (\ref{eq:selfcs2}) can be further restricted to a limited set (leaving out the states close to band edges) to achieve a desired accuracy.

Motivated by these findings, we simplify the PNS calculations by limiting progressively smaller number of total states contributing to pairing. The corresponding output of $\Delta_{\rm OP}$, as its percentage deviation from the GIMT value, is shown in Fig.~(\ref{fig:width}d) against the fraction of normal states participated in pairing. To illustrate our choice of restricted states for $V=2.5$, we show the bounding box $BH$ in Fig.~(\ref{fig:width}c) by a thin dotted line that includes about 19\% of the normal states for pairing, and results into more than 99\% accuracy in $\Delta_{\rm OP}$ (the last data point along $x$-axis in Fig.~(\ref{fig:width}d)). It is apparent that our bounding box encloses states that subscribe to $|\Delta_{\alpha \beta}|$ of significance~\footnote{Accuracy of results could be enhanced by fine tuning our choice of bounding box, but we primarily focussed on a simple and robust scheme.}.
Evidently, PNS results achieve perfection when increasing fraction of states are included. Yet, we see that only about $10\%$ of ${\rm NS}_{\rm GIMT}$  ensures $95\%$ accuracy of the results, even for disorder as large as $V=2.5$!

\subsection{Pairing theory with `uncorrelated' NS and with a different model of disorder}\label{sec:conc}

We discuss below the prospects of our PNS proposal in terms of `uncorrelated' normal states, in which all Gutzwiller factors are set to unity. The impressive match of $\Delta_{\rm OP}$ from such pairing theory using ${\rm NS}_{\rm IMT}$, in comparison with the plain BdG results has already been analysed in Sec.~(\ref{sec:odlro}). In fact, we found that the IMT-normal states are very close to the original `exact eigenstates', except, of course, for the Hartree- and Fock-shifts, which adds only weak corrections in the absence of Gutzwiler renormalization. While the success of PNS formalism is evident, there are practical concerns for the applicability of such implementation.
The ${\rm NS}_{\rm IMT}$ are naturally incapable of accounting for the strong correlation effects, crucial for the qualitative physics of the strongly correlated superconductors. In addition, the pairing of ${\rm NS}_{\rm IMT}$ misses the near-diagonal nature of $|\Delta_{\alpha \beta}|$ as found in Fig.~(\ref{fig:width}) for ${\rm NS}_{\rm GIMT}$, making the ${\rm NS}_{\rm IMT}$ less useful, for deriving technical advantages over IMT calculations.

We also verified that the results and conclusions of PNS formalism remain valid even with a model of `concentration impurity', in which $n_{\rm imp}$ fraction of the (random) lattice sites contain a fixed disorder potential $V_0$, provided we use $V_0 \le 3$. Stronger $V_0$ brings in subtle effects even in GIMT implementation~\cite{PhysRevB.95.014516}.

\section{Discussions}\label{sec:discuss}

The impressive match between the PNS and GIMT results is inspiring from the perspective of developing simple understanding on the complex physics of disordered and strongly correlated superconductors. However, we believe that it is the conceptual advances offered by PNS technique, as described in the previous sections, which have far reaching values. We will discuss below a crucial notional gains from the PNS proposal.

\subsection{Insensitivity of inhomogeneity in pairing}\label{subsec:inhom}

Our results make it evident that inhomogeneities are less relevant for pairing in case of strongly correlated dSC. This has already been illustrated in Ref.~~\onlinecite{1367-2630-16-10-103018}, by matching the spectral density of states evaluated in GIMT for $V \le 3t$, with its d-wave BCS form convoluted with the near-Gaussian GIMT distribution of $\Delta_{ij}$. Here, we argue for a more direct evidence to this assertion by noting that the spatial inhomogeneities in the Gutzwiller factor $g^{xy}_{ij}$, arising from the spatial fluctuations in the local density, has little role in the final self-consistent output of $\Delta_{ij}$ on the bonds. This is, however, only true, provided that the {\it correct} ${\rm NS}_{\rm GIMT}$ is obtained by taking care of all inhomogeneities in their construction. For concreteness, we can consider three independent implementations of $g^{xy}_{ij}$, with a progressive degree of approximations of the inhomogeneities: (a) A full self-consistency in local density $\rho_i$ in the definition of $g^{xy}_{ij}$ is achieved during the iterative update of $\Delta_{ij}$ during the pairing stage following Eq.~(\ref{eq:deltadef}). (b) We fix the inhomogeneous density profile to its form as obtained in ${\rm NS}_{\rm GIMT}$, without any update during the pairing self-consistency. (c) In the extreme approximation, we set $g^{xy}_{ij}=(1-0.5\rho)^{-2}$ for the purpose of pairing self-consistency.
Obviously, each degree of approximation is associated with significant computational gains. We find that even with the most drastic approximation, the resulting order parameters are in good agreement (within 10\%) with the GIMT findings. On the other hand, we found that an approximate handling of heterogeneities in the normal state leads to significant deviation of the final results.

\subsection{What makes d-wave anisotropy of pairing so robust?}\label{subsec:dwave}

\begin{figure}[t]
\centering
  \begin{tabular}{@{}cc@{}}
    \includegraphics[width=.23\textwidth]{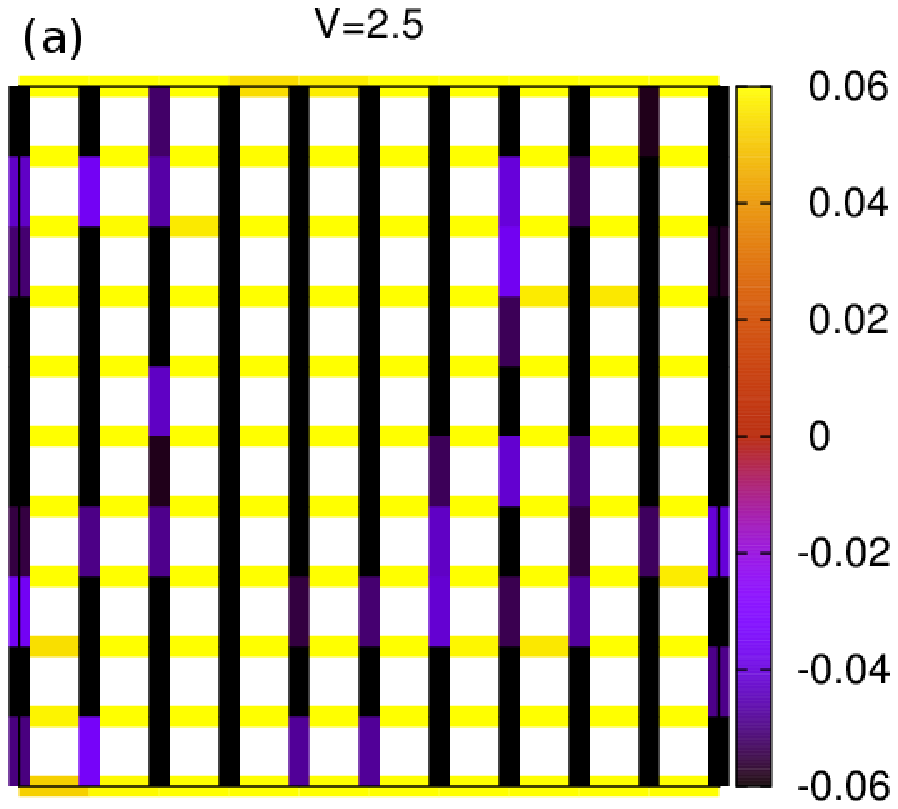} &
    \includegraphics[width=.23\textwidth]{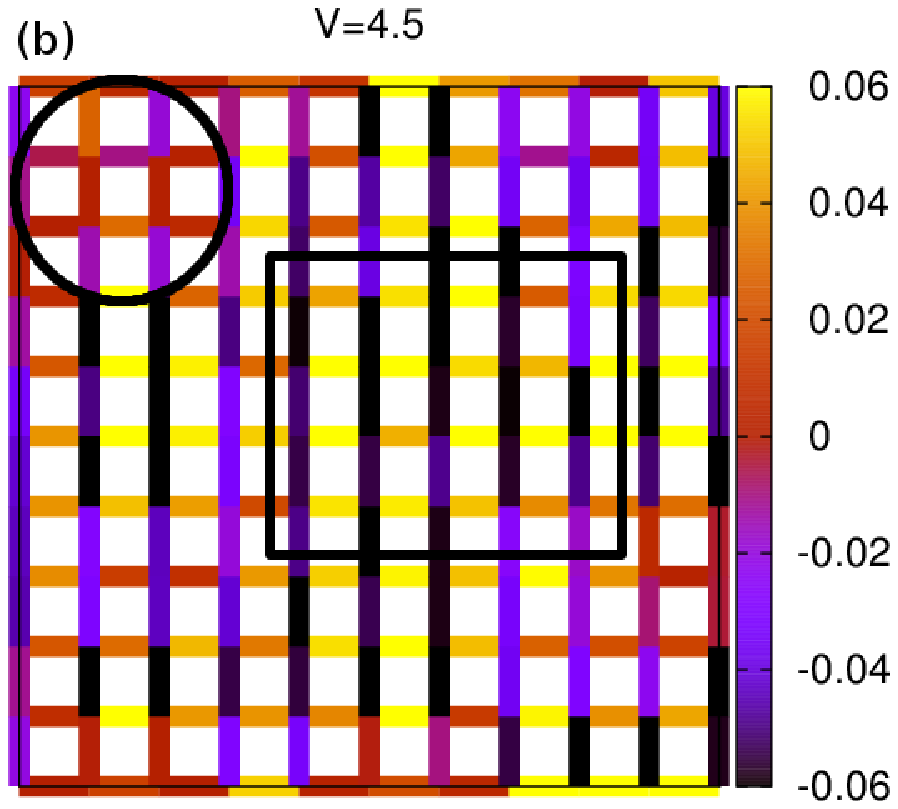} 
  \end{tabular}
  \caption{$\Delta_{ij}$ on each bond for a section of the lattice for: (a) $V=2.5$, and (b) $V=4.5$ for a specific realization of disorder. The $\pi/2$ phase difference between $\Delta_{i,i+\hat{x}}$ and $\Delta_{i,i+\hat{y}}$ survives over the entire lattice as seen in (a). The larger disorder strength of panel (b) still supports the d-wave anisotropy in most parts (highlighted by the square boundary). It also features regions of strong potential fluctuations (marked by circular boundary), where $\Delta_{i,i+\hat{x}}$ and $\Delta_{i,i+\hat{y}}$ are closer in magnitude, but only when both are vanishingly small!}
\label{fig:Bonds}
\end{figure}

Why does not AG-theory capture the insensitivity of strongly correlated d-wave superconductors to impurities? Admittedly, such strongly coupled systems with short coherence length $\xi$, fall outside the scope of a true AG description. However, our PNS formalism offers a simple and intuitive perspective for the distinct outcome of the GIMT findings. Such results (or the results from PNS, which produces essentially identical results as GIMT) of the spatial profile of $\Delta_{ij}$ on each bond on a square lattice is shown in Fig. (\ref{fig:Bonds}a), for a specific realization of disorder. 

We witness pairing amplitudes of opposite signs but of nearly equal strengths on bonds along $\hat{x}$- and $\hat{y}$-directions from each site for $V=2.5$ (See Fig.~(\ref{fig:Bonds}a)). Such a phase differences of $\pi/2$ between adjacent orthogonal bonds is the hallmark of its $d_{\rm x^2-y^2}$ anisotropy of pairing amplitude~\cite{PhysRevLett.71.2134,PhysRevB.54.R9678,PhysRevLett.73.593} in the clean systems, and remains near-perfect even at $V=2.5$! With the introduction of disorder, AG theory predicts that the impurity scattering `mixes-up' such sensitive phase relations, and thereby depletes d-wave superconductivity~\cite{AGtheory}.
Instead, we find a healthy d-wave anisotropy to survive.
But, this is naturally expected within the PNS formalism -- there is no disorder left to scramble phases at the second stage of `pairing', they are all consumed in generating the normal states at the first stage of calculations!

Do such phase relations continue to hold for stronger disorders? While additional considerations are necessary for pushing the applicability of the PNS method to larger $V$, an extension of GIMT-type calculation upon including localization physics for $V \ge 3$ has already been reported in Ref.~~\onlinecite{PhysRevB.95.014516}, and those results offer a significant pointer. By ramping up $V$ in such calculations, we find that for $V=4.5$, the local pairing amplitude tends to zero identically on {\it both} $\hat{x}$- and $\hat{y}$-bonds in regions of strong fluctuation of disorder potential (marked by circular boundary in Fig.~(\ref{fig:Bonds}b)). Yet, the d-wave anisotropy remains intact in regions possessing a healthy $\Delta_{ij}$ (marked by square boundary), albeit some inhomogeneity. Thus, impurities can affect superconductivity by locally collapsing the self-consistent pairing amplitudes, which are due to the localization properties of the normal states, but are not because of scrambling of the d-wave anisotropy.

\section{Conclusion}\label{sec:conclu}

In conclusion, we presented a description of disordered and strongly correlated d-wave superconductors by implementing simple pairing ideas of Anderson, but extending it by including the effects of strong electronic correlations as well as disorder induced inhomogeneities.
The impressive match of the results from the proposed PNS method and GIMT findings is encouraging. In addition to offering a deeper understanding of the GIMT findings, our formalism sheds important light on some shortcomings of the conventional wisdom. The pivotal advance offered by the PNS formalism lies in identifying the underlying effective one-particle states that participate in Cooper-pairing in unconventional superconductors. This motivates future survey of the properties of ${\rm NS}_{\rm GIMT}$ by probing them using various means, and in particular on their temperature dependences. It will also be interesting to consider the robustness of the PNS formalism upon including the physics of `competing orders' in ${\rm NS}_{\rm GIMT}$ and their role in subsequent Cooper-pairing. 

\begin{acknowledgments}
We thank Indranil Paul and Kazumasa Miyake for useful discussions.
\end{acknowledgments}

\bibliographystyle{apsrev}
\bibliography{Draft.bib}

\end{document}